\documentclass[a4paper,10pt,twoside]{cpc-hepnp}

\usepackage{multicol}
\usepackage{graphicx}
\usepackage{booktabs}
\usepackage{amssymb,bm,mathrsfs,bbm,amscd}
\usepackage[tbtags]{amsmath}
\usepackage{lastpage}
\usepackage{CJK}
\usepackage{float}
\usepackage{graphicx}
\usepackage{natbib}

\begin{document}

\begin{CJK*}{GBK}{song}



\title{Development of a sub-milimeter position sensitive gas detector \thanks{Supported by NSFC (11075095) and Shandong Province Science Foundation (ZR2010AM015) }}

\author{%
      DU Yanyan$^{1}$
\quad XU Tongye$^{1}$
\quad SHAO Ruobin$^{1}$\\
\quad WANG Xu$^{1}$
\quad ZHU Chengguang$^{1;1)}$\email{zhucg@sdu.edu.cn}%
}
\maketitle

\address{%
$^1$ MOE key lab on particle physics and particle irradiation,\\Shandong University, Ji'nan 250100, China
}

\begin{abstract}
A position sensitive thin gap chamber has been developed. The position resolution was measured using the cosmic muons. This paper presents the structure of this detector, position resolution measurement method and results.
\end{abstract}




\begin{multicols}{2}
\section{Introduction}
 TGC( Thin Gap Chamber) used in ATLAS experiment~\citep{ATLAS1997ad} shows good performance in the fast response and time resolution, but with limited position resolution. The improvement of the position resolution with the timing performance retained is straightforward for its flexibility to be used in the future experiments and radiation measurement, for example, the upgrade of the trigger system of ATLAS experiment. The main goal of the study described in this paper aimed to build a prototype detector based on TGC, which can have a position resolution better than 300${\mu}m$, while keeping timing performance not deteriorated

  The TGC detector operates in saturated mode by using a highly quenching gas mixture of carbon dioxide and n-pentane, $55\%$ :$45\%$, which has many advantages, such as small sensitivity to mechanical deformations, small parallax, small Landau tails and good time resolution, but a position sensitivity of around 1$cm$, decided by the geometrical width of the readout channel and the strength of the induced signal. To improve the position resolution, we concentrated on the improving of the method of signal readout by fine tuning of the structure of the detector.

 The new detector, named as pTGC(precision Thin Gap Chamber) based on the ATLAS TGC, is constructed and tested. We found the position resolution can be  improved to be less than $300{\mu}m$, which meets the requirements.

In section 2, the structure of pTGC detector is described. Section 3 is devoted to pTGC's position resolution measurement. Results of the measurement is summarized in section 4.

\begin{figure*}

\begin{minipage}[t]{0.5\linewidth}

\centering

\includegraphics[width=7cm]{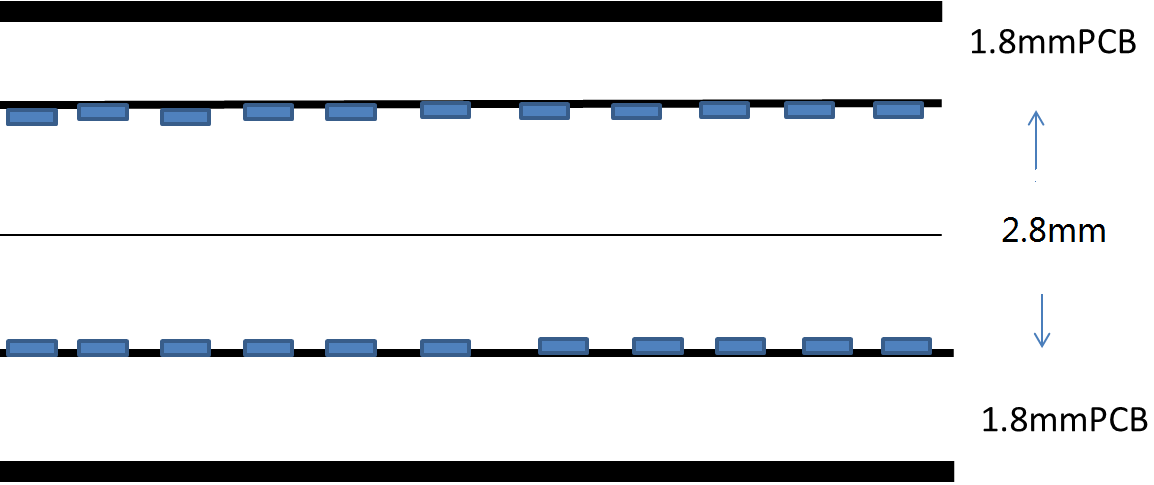}

\caption{The schematic structure of pTGC-I chamber. Anode wires are placed in the middle, with copper strip etched on the inner surface of the PCB board, perpendicular to the wire direction. }

\label{fig:side:a}

\end{minipage}%
\hspace{3ex}
\begin{minipage}[t]{0.5\linewidth}

\centering

\includegraphics[width=7.8cm]{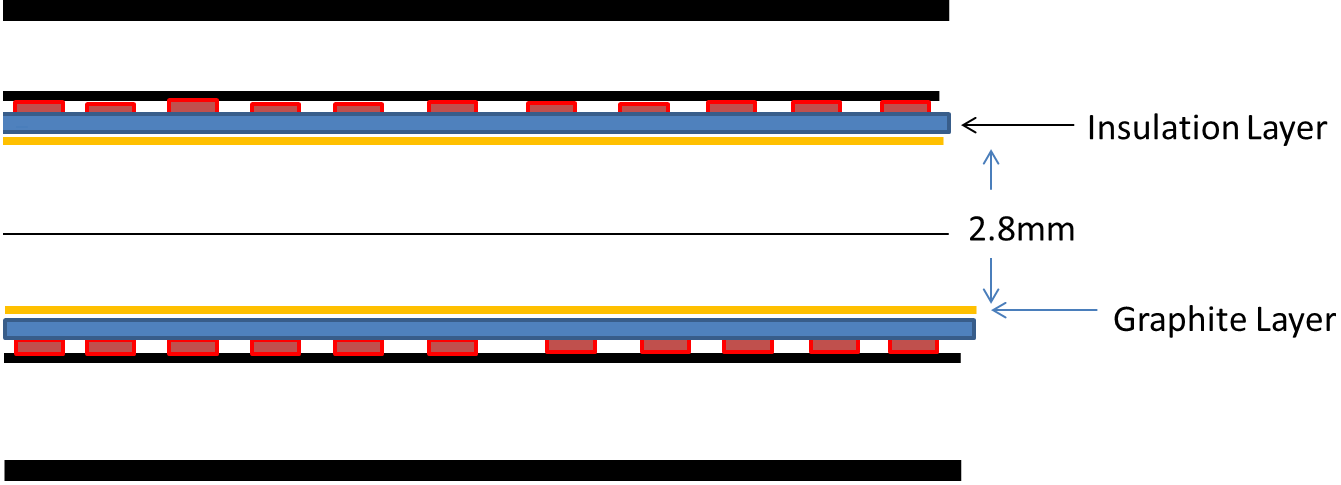}

\caption{The schematic structure of pTGC-II chamber. Compared to pTGC-I, additional isolation layer and graphite layer cover the etched copper strips.}

\label{fig:side:b}

\end{minipage}

\end{figure*}

\section{Construction of pTGC}
In the pTGC development, two versions of detectors are constructed and tested, which are referred as $pTGC-I$ in the first stage and $pTGC-II$ in the second stage, respectively.

The schematic structure of $pTGC-I$ is shown in Fig.~\ref{fig:side:a}, similar to the structure of ATLAS TGC, except that the position of the strips for signal collection are modified. 48 copper strips of $0.8mm$ wide and $0.2mm$ spaced are etched on the inner surface of the 2 parallel PCB boards, which form a thin spaced chamber. The wires, segmented at $1.8mm$ interval and perpendicular to the strip direction, are sandwiched in between the two PCB boards. The resulted size of the detector is defined by the number of wires and strips, which are $290mm{\times}50mm$.

In the test of $pTGC-I$, the discharge happened between wires and strips resulted in fatal damage on the frontend electronics, even though we have designed a protection circuit to insert between detector and frontend electronics board. This means an instability for big detector and for long time running. Besides, the induced charge on strips spread roughly $5$ to $6mm$, which leaves limited rooms for reducing the quantity of the channels by enlarging the width of strips. Based on $pTGC-I$, the $pTGC-II$ is developed to deal with these problems.

The schematic structure of $pTGC-II$ is shown in the Fig.~\ref{fig:side:b}. The strip width is enlarged to $3.8mm$ ($0.2mm$ spaced), and a thin ($100{\mu}m$) insolation layer is pasted on the strip layer. The isolation layer is then coated with a thin (~$30{\mu}m$) graphite layer as the electric ground to form the electric field with wires. This graphite layer acts as the protector of the frontend electronics from discharge and can enlarge the spreading size of the induced charge on the strip layer. We tune the resistivity of the graphite layer to be around $100k\Omega$, considering the diffusion speed of the charge, as well. the resulted size of $pTGC-II$ is $290mm*200mm$.

Both detector use gas mixture of carbon dioxide and n-pentane, $55\%$ :$45\%$,  as working gas, and the anode wire is set to high voltage of 2900v, which are all the same configuration as the ATLAS TGC detector to maintain the its features relative to the time measurement of the detector.

\section{Position resolution measurement}
\begin{figure}[H]
   \begin{minipage}{0.45\textwidth}
    \includegraphics[height=5.3cm, width=7.4cm, clip,scale=0.4]{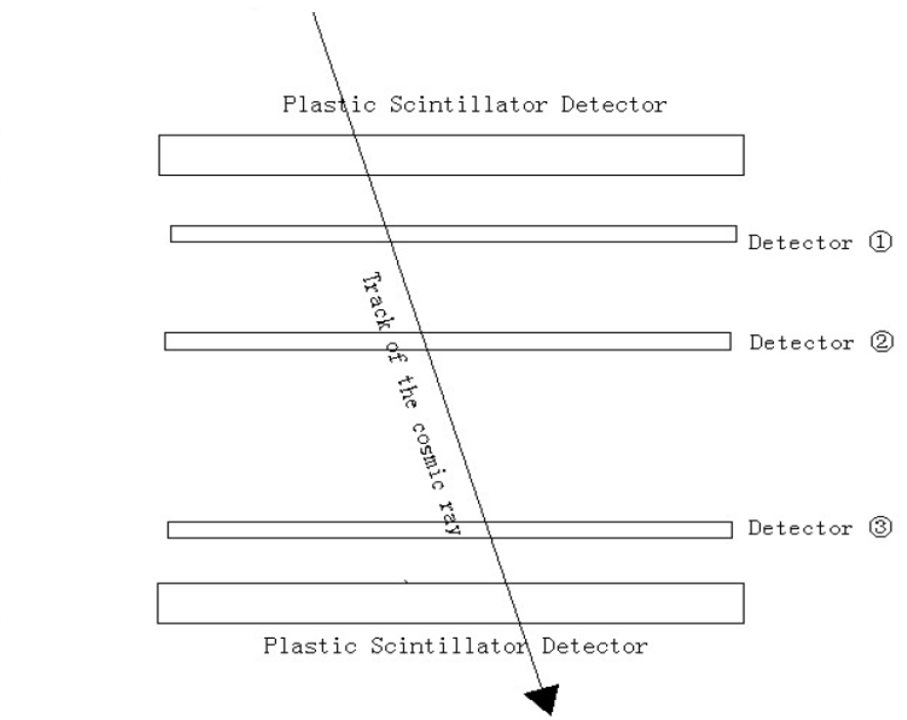}
    \caption{The comic muon hodoscope used for the chamber testing. Plastic scintillator detector are used for trigger. 3 identical pTGC chambers placed in parallel in between the 2 scintillator detectors.}
    \label{fig:detectorSetting}
    \end{minipage}
\end{figure}

With 3 layers of identical chambers placed in parallel, and 2 layers of scintillator detectors to build a muon hodoscope(see Fig~\ref{fig:detectorSetting}), the $pTGC-I$ and $pTGC-II$ detectors are tested. The induced charge on each strip is integrated for the the position calculation based on the charge center-of-gravity algorithms. The measured hitting position on the 3 layers of chamber are supposed to be aligned into a straight line concerning the penetration power of muons. The residue of the position relative to the straight line is then used to calculate the position resolution of the detectors.

\subsection{Signal definition}
\begin{figure}[H]
   \begin{minipage}{0.45\textwidth}
    \includegraphics[height=4.5cm, width=7.4cm,clip, scale=0.4]{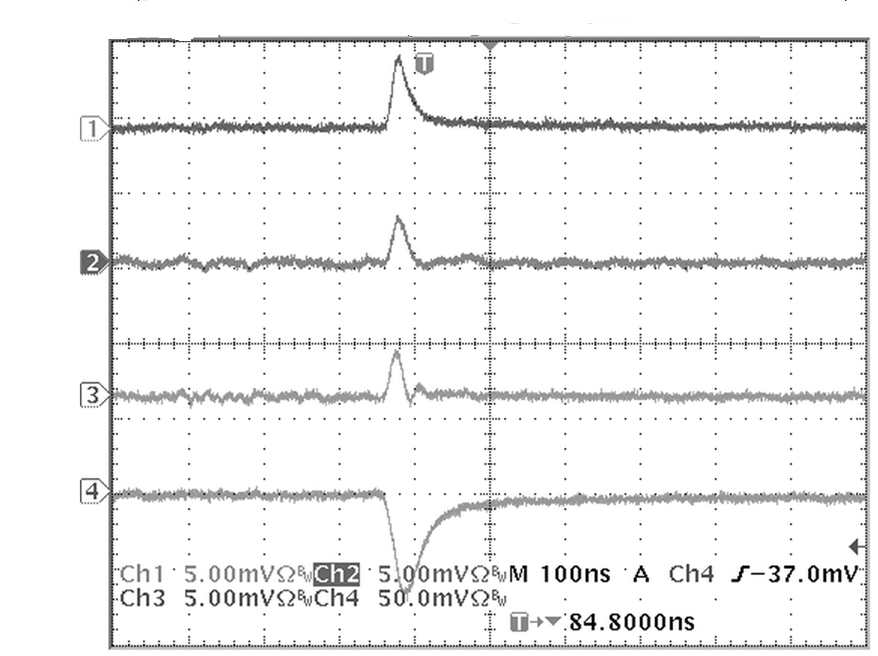}
    \caption{The observed signals on wires and several copper strips induced by the same incident cosmic ray. The signal on wire are negative, and the signal on strips are positive.}
    \label{fig:anasignal}
    \end{minipage}
\end{figure}

Using oscilloscope, we first observed the induced signal in one wire group and 3 adjacent strips (limited by channels of oscilloscope), as shown in Fig.~\ref{fig:anasignal}. It's apparent that the signals are great significant above the noise and the signals on the strip are in different magnitudes as expected.

For position resolution measurement, we designed a much more complicated DAQ(data acquisition) system based on gassiplex frontend electronics~\citep{minghui} to readout and digitize the induced charge from a quantity of channels of the 3 chambers in a more complex hodoscope~\citep{tongye}. Once the two scintillator detector of the hodoscope are both fired, the DAQ is triggered. The trigger signal is sent to the detector front end electronics, which then close the gate for the discharge of capacitance which has integrated the signal charge on. The charge on the capacitance are then read out one by one controlled by the clock distributed from the DAQ system. The charge are then digitized and saved into computer.

The digitized charge, denoted by $Q_i$ where $i$ is the channel number, consists of three parts: electronic pedestal, noise, and charge induced by muon hit.

\begin{figure}[H]
  \begin{minipage}{0.25\textwidth}
    \includegraphics[width=7.4cm, height=4.5cm, clip,scale=0.4]{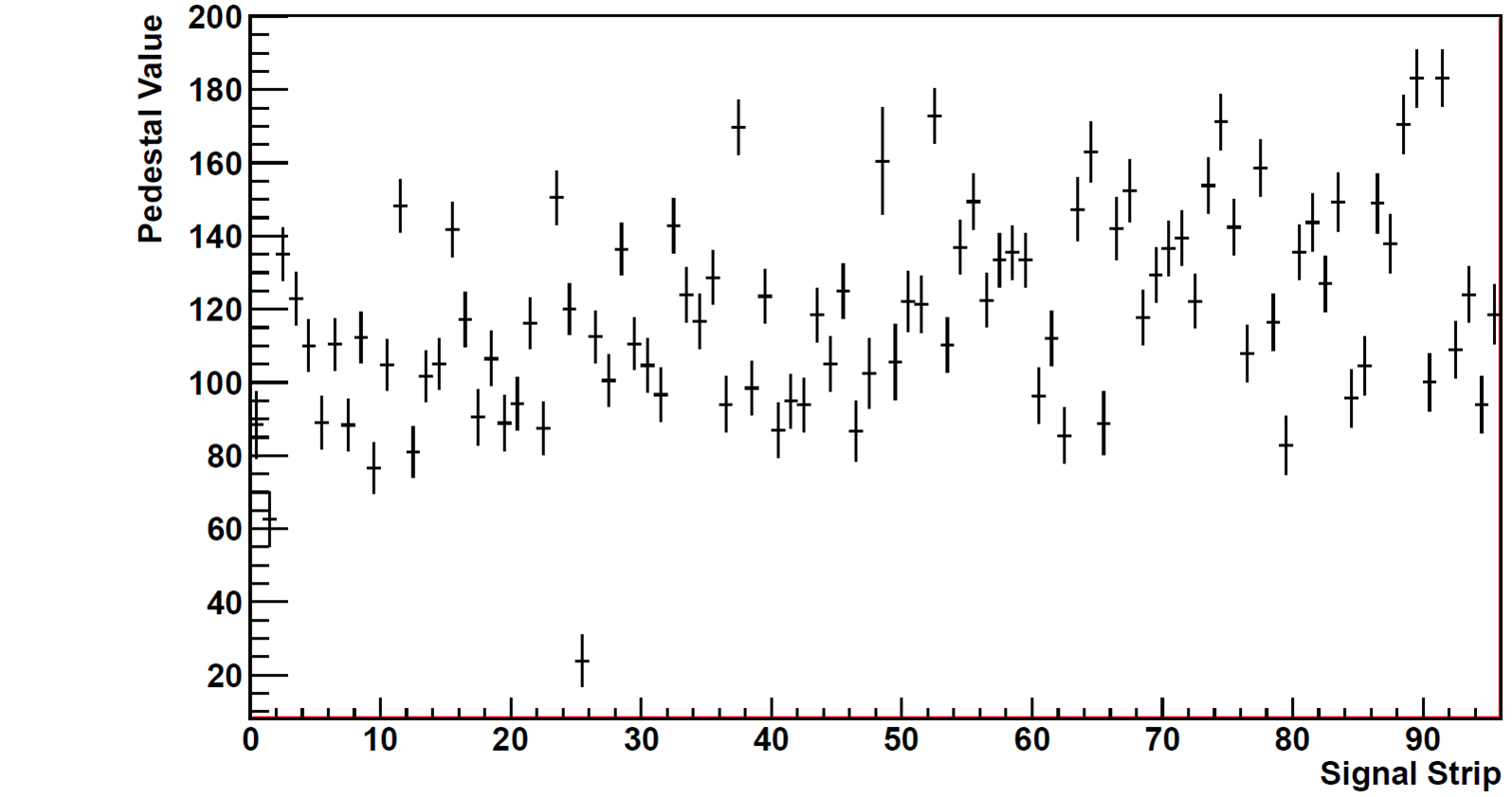}
  \end{minipage}
  \caption{\small\sf The noise and pedestal distribution of 96 signal strips in one chamber (The x-axis is the signal strip number, the vertical coordinate is the pedestal value and the error bars presents the noise of that channel.) }
  \label{fig:detectorProperty}
\end{figure}

First of all, we need to figure out the pedestal and noise for each channel. The method is to histogram the integrated charge for each channel using a soft trigger where no real muon induced signal appear in the data. Fitting the histogram with a gaussian function to get the pedestal and the noise, denoted by $P_i$ and $\sigma_i$, as shown in Fig.~\ref{fig:detectorProperty}, where the height of the histogram represents the pedestal and the error bar represents the noise of that channel.

In the analysis, if $Q_i>P_i+3\sigma_i$, the channel is considered to be fired by real muon hit, and the signal charge is calculated as:
\begin{equation} \label{equ:timeDiff}
S_i=Q_i-P_i,
\end{equation}

\begin{figure}[H]
  \begin{minipage}{0.25\textwidth}
    \includegraphics[height=4.4cm,width=7.5cm,clip,scale=0.4]{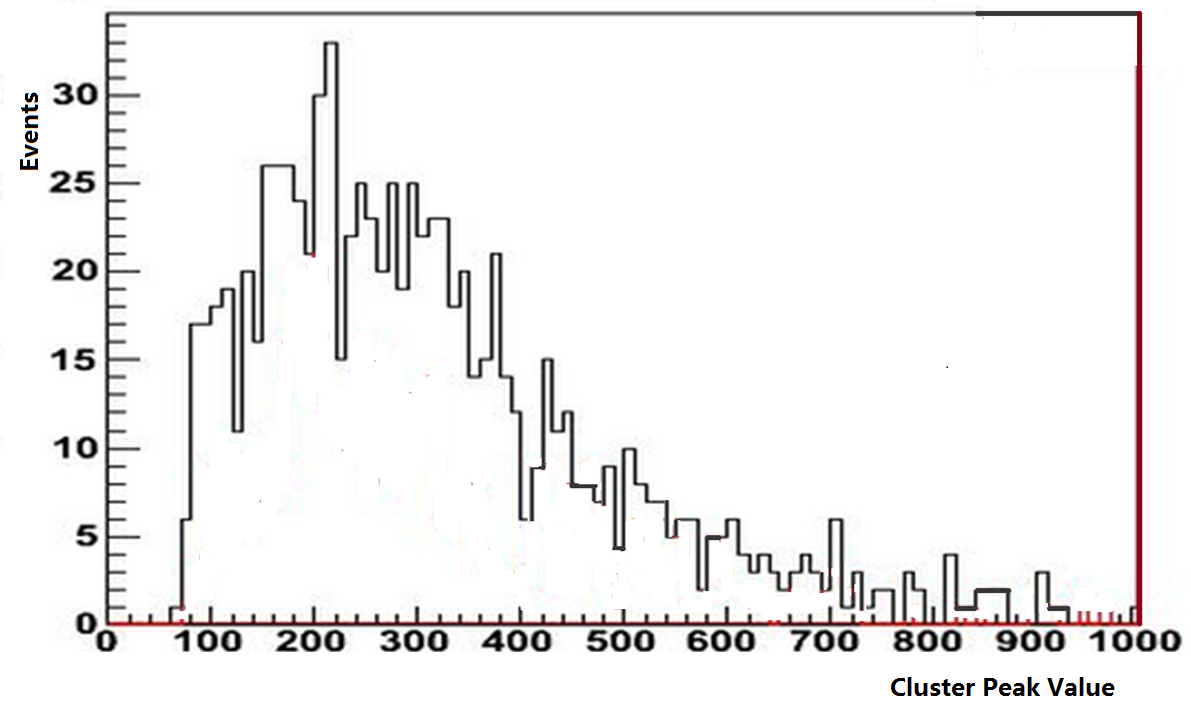}
  \end{minipage}
  \caption{\small\sf The distribution of the largest signal in one cluster. The x-axis is the digitized charge collected. }
  \label{fig:peaksignal}
\end{figure}

\begin{figure}[H]
  \begin{minipage}{0.25\textwidth}
    \includegraphics[height=4.4cm, width=7.5cm,clip,scale=0.4]{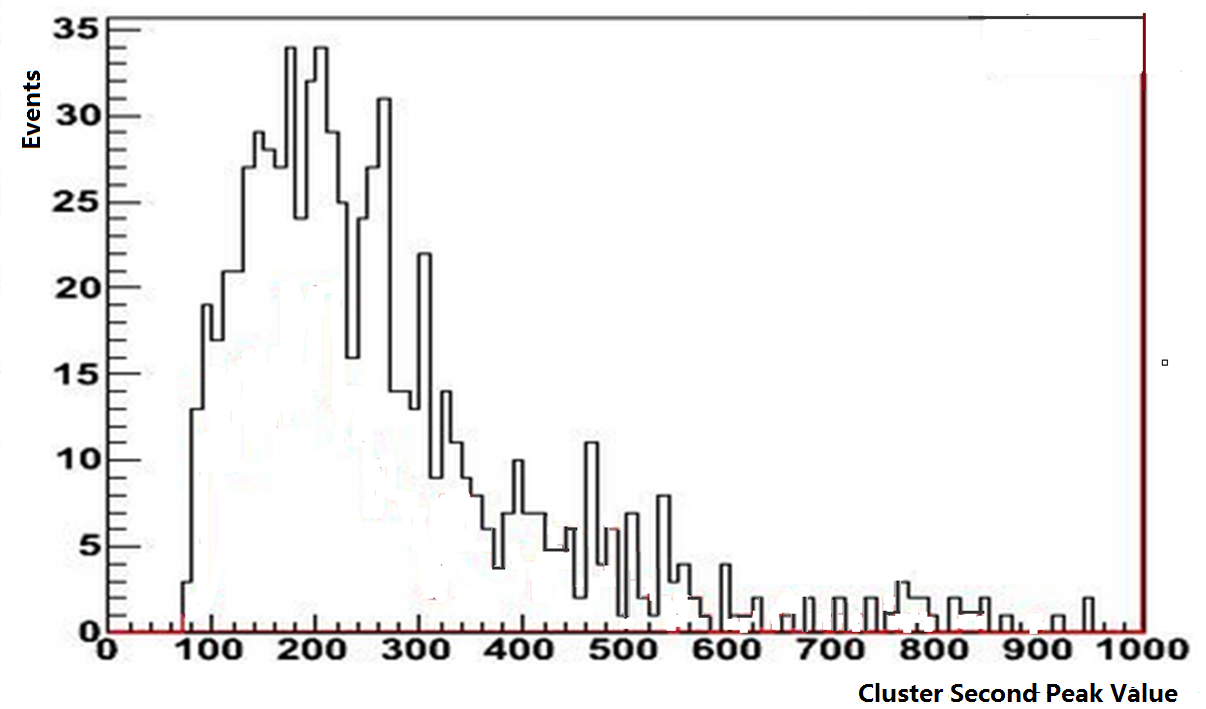}
  \end{minipage}
  \caption{\small\sf The distribution of the second largest signal in one cluster }
  \label{fig:seconpeak}
\end{figure}

\begin{figure}[H]
  \begin{minipage}{0.25\textwidth}
    \includegraphics[height=4.4cm,width=7.5cm,clip,scale=0.4]{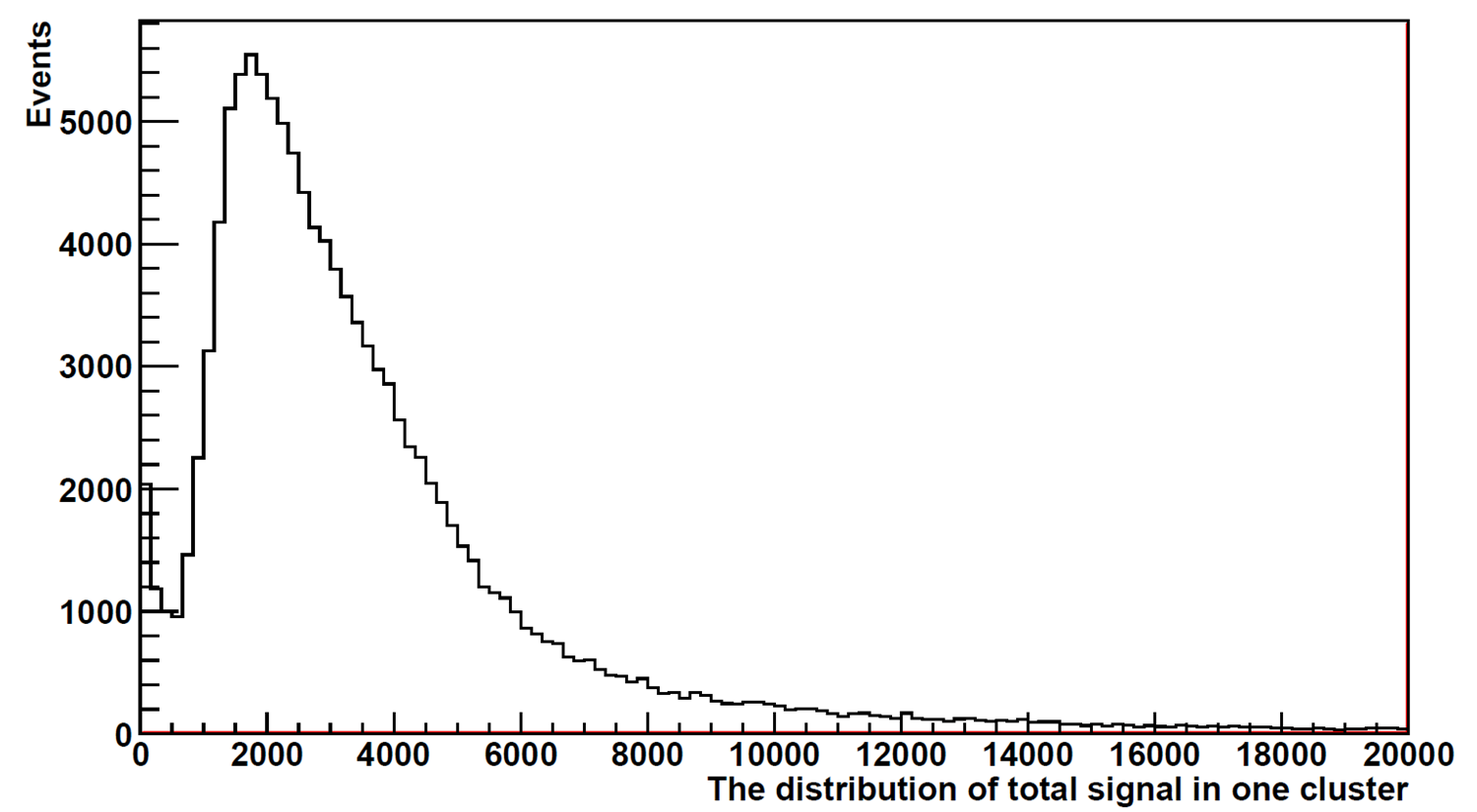}
  \end{minipage}
  \caption{\small\sf The distribution of total charge induced in one cluster }
  \label{fig:totalcharge}
\end{figure}

The signal magnitude distribution of the largest signal in each cluster (cluster definition is in next section), named as the peak signal, is shown in Fig.~\ref{fig:peaksignal}. The distribution of the second largest signal in each cluster, named as second peak signal, is shown in Fig.~\ref{fig:seconpeak}. The distribution of the sum of all charge in one cluster is shown in Fig.~\ref{fig:totalcharge}. The correspondence between the magnitude of the signal and the charge is $1fC/3.6bits$. We can then calculate that the maximum probable charge of the largest signal in one cluster is $69fC$, the maximum probable total charge of one cluster is $470fC$, which is consistent with the measurement in ~\citep{ATLAS1997ad}

\subsection{Cluster definition}
\label{cluster}
\begin{figure}[H]
  \begin{minipage}{\columnwidth}
    \includegraphics[height=4cm,width=8cm, clip]{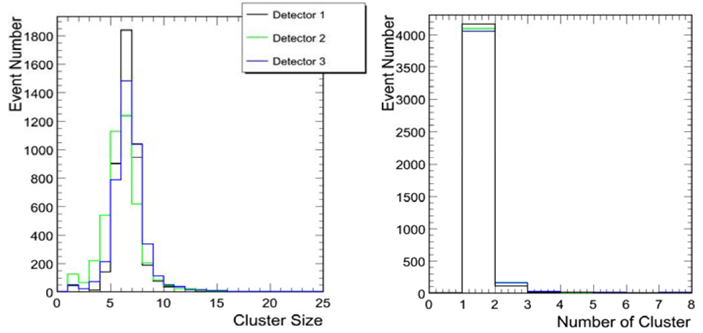}
  \end{minipage}
  \caption{\small\sf For pTGC-I: (Left) The distribution of cluster size (quantity of strips in one cluster). (Right) The quantity of cluster in one chamber per triggered event. }
  \label{fig:FclusterProperty}
\end{figure}

\begin{figure}[H]
  \begin{minipage}{\columnwidth}
    \includegraphics[width=8cm,height=4cm, clip]{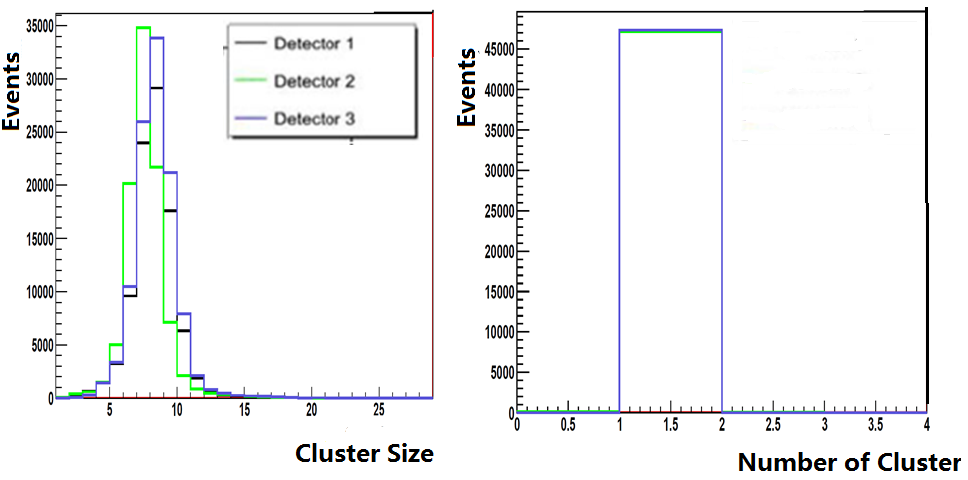}
  \end{minipage}
  \caption{\small\sf For pTGC-II: (Left) The distribution of cluster size (quantity of strips in one cluster). (Right) The quantity of cluster in one chamber per triggered event.}
  \label{fig:SclusterProperty}
\end{figure}

 The induced charge by the incident muons are distributed on several adjacent strips, which are grouped in "cluster" in the analysis and used for the hit position calculation. In one event, we search all the channels of one detector, and define group of fired adjacent strips without space as a cluster. To suppress the fake signals from noise, if the cluster contains only one strip, the cluster is dropped. The cluster size and number of cluster per detector per event are shown in Fig.~\ref{fig:FclusterProperty} for $pTGC-I$ and Fig.~\ref{fig:SclusterProperty} for $pTGC-II$. It can be seen that in both cases one cluster contain average six strips and almost every event contains one cluster, which is consistent with the expected. The hit position is then calculated for each cluster by
\begin{equation} \label{equ:clusterCoordinate}
x=\sum_{i}(S_{i}*x_{i})/\sum_{i}(S_{i}),
\end{equation}
where $x_i$ is the center coordination of the $i-th$ strip.

\subsection{Position resolution}
As redundant design, the strips are etched on both inner surface of the PCB boards. Signals will be induced by the same avalanche on the 2 face-to-face strips, which corresponds to an double measurements of a single hit. To compare the two measurements, denoted as $x_1$ and $x_1'$, we fill $x_1-x_1'$ into histogram to see the broadness of the distribution. From a simple gaussian function fit, we observed a narrow width of around $36{\mu}m$, which means that the electronics noise effect on the resolution is much small. This is consistent with the expectation when to compare Fig.~\ref{fig:detectorProperty} and Fig.~\ref{fig:peaksignal}, where it shows the signal is great significant compared to the noise.

\begin{figure}[H]
  \begin{minipage}{0.25\textwidth}
    \includegraphics[width=8.1cm,height=3.9cm,clip,scale=0.4]{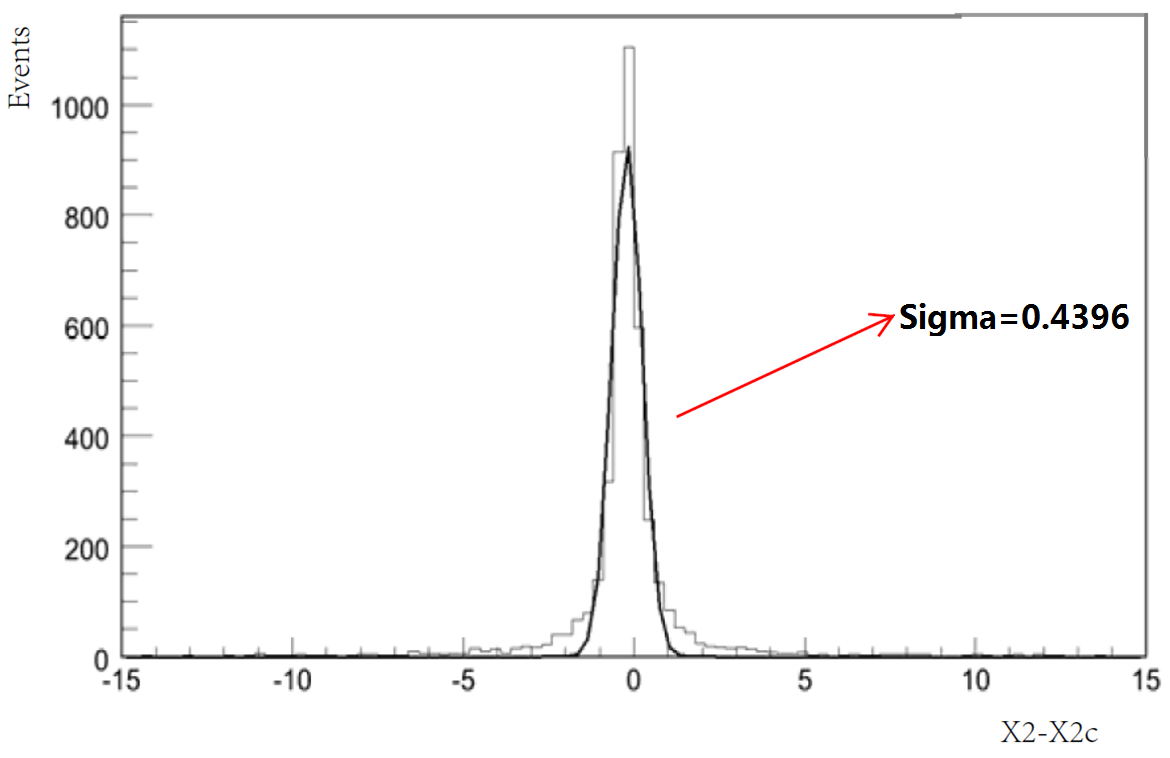}
  \end{minipage}
  \caption{The distribution of $x_{2}-x_{2c}$ for pTGC-I. The corresponding position resolution of the chamber is $\sigma=\frac{w}{\sqrt{1+k^2}}=\frac{439{\mu}m}{1.22}=359{\mu}m$. }
  \label{fig:oldResolution}
\end{figure}

\begin{figure}[H]
  \begin{minipage}{0.25\textwidth}
    \includegraphics[height=3.4cm, width=8cm, clip,scale=0.4]{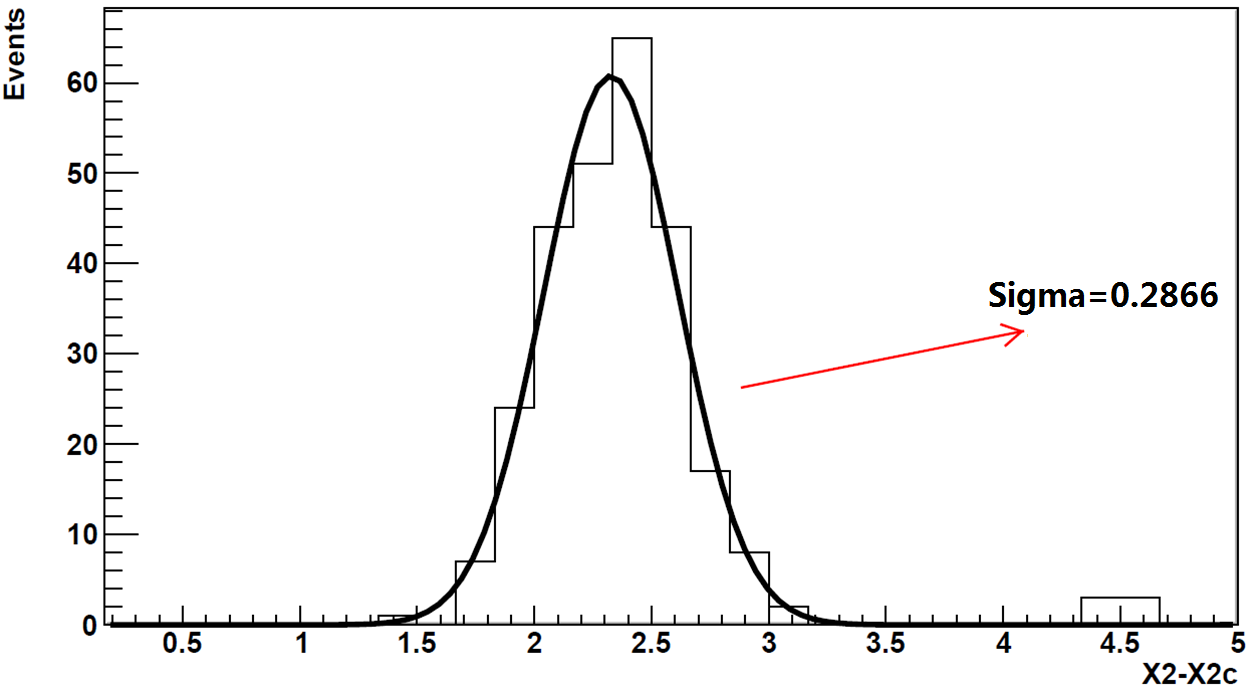}
  \end{minipage}
  \caption{ The distribution of $x_{2}-x_{2c}$ for pTGC-II. The corresponding position resolution of the chamber is $\sigma=\frac{w}{\sqrt{1+k^2}}=\frac{286{\mu}m}{1.22}=233{\mu}m$.}
  \label{fig:updatingResolution}
\end{figure}

After the three hit positions $x_{1}$, $x_{2}$, $x_{3}$ are calculated for the 3 parallel chambers, to simplify the calculation, we first use $x_{1}$ and  $x_{3}$ to calculate the expected hit position on the second layer $x_{2c}$:
\begin{equation} \label{equ:x2c}
x_{2c}=x_1\frac{L_{23}}{L_{12}+L_{23}}+x_3\frac{L_{12}}{L_{12}+L_{23}},
\end{equation}
where $L_{12}$ and $L_{23}$ are the vertical distance between the detector 1,2 and 2,3.
To assume the same position resolution $\sigma$ for the 3 identical detectors, we know the resolution of $x_{2c}$, with the error propagation, is:
\begin{equation} \label{equ:x2csigma}
\sigma_{2c}=\sqrt{\frac{L_{23}^2}{(L_{12}+L_{23})^2}+\frac{L_{12}^2}{(L_{12}+L_{23})^2}}   \sigma{\equiv}k\sigma,
\end{equation}
 Filling $x_{2}-x_{2c}$ into the histogram and then fit with gaussian function, the width is $w=\sqrt{1+k^2}\sigma$. So we can directly calculate the position resolution of the detector as
 \begin{equation} \label{equ:sigmac}
\sigma=\frac{w}{\sqrt{1+k^2}}.
\end{equation}
 From Fig.~\ref{fig:oldResolution} and Fig.~\ref{fig:updatingResolution}, we can obtain that the position resolution are $359um$ for $pTGC-I$ and $233um$ for $pTGC-II$. In both of the cases, the detector resolution has reach our design requirement. In test, we see that $pTGC-II$ are more stable with the graphite layer protection and achieve a better resolution even with less channels.


\begin{figure}[H]
  \begin{minipage}{0.25\textwidth}
    \includegraphics[height=4.2cm,width=6.7cm, clip,scale=0.4]{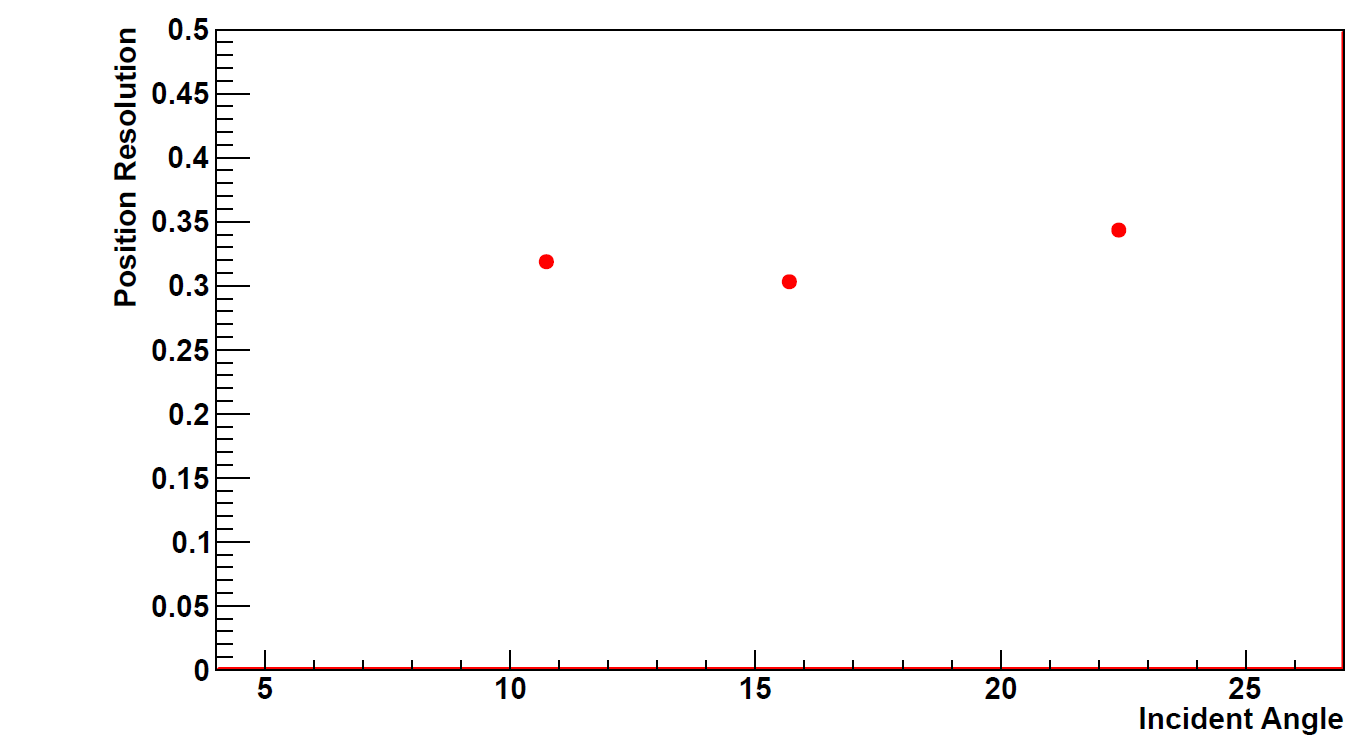}
  \end{minipage}
  \caption{\small\sf The position resolution variance relative to the incident angle of the cosmic rays. The x-axis is the incident angle of cosmic rays.  }
  \label{fig:angles}
\end{figure}

To look at the dependence of the position sensitivity of the detector to the incident angle of the muon, we divide the data into groups. Each group of data contains the events of muon with specific incident angle. To redo the analysis above, the result is shown in Fig.~\ref{fig:angles}, which shows that the position resolution of $pTGC-II$ is insensitive to the incident angle of muons.

\begin{figure}[H]
  \begin{minipage}{0.25\textwidth}
    \includegraphics[height=4.2cm,width=6.7cm, clip,scale=0.4]{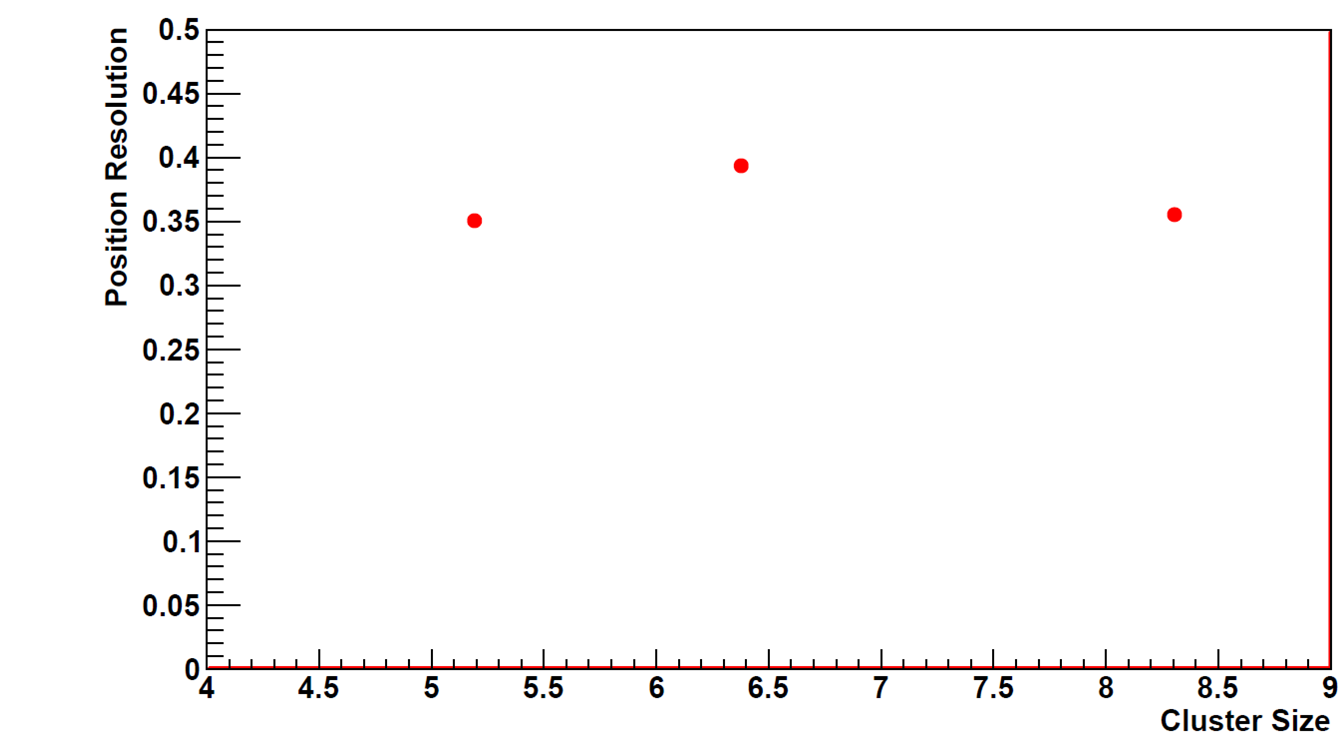}
  \end{minipage}
  \caption{\small\sf The position resolution variance relative to the quantity of strips in one cluster used for position calculation. The x-axis is the quantity of strips in one cluster used for position calculation. }
  \label{fig:lesstrip}
\end{figure}

 To check the effect of the electronic noise, we use part of the top highest signals in one cluster to calculate the position resolution. The result is shown in Fig.~\ref{fig:lesstrip}, which shows that the resolution are similar and the electronic noise doesn't affect much.

\section{Summary}
Two pTGC version $pTGC-I$ and $pTGC-II$, have been constructed and tested. With the basic structure and working gas unchanged, the detector can attains the exiting features like good time resolution and fast response, which are essential for trigger. By revising the signal collecting structure and method, the position resolution is improved from the level of centimeter to be less than $300{\mu}m$, which meet the requirement of design. To be noticed that the resolution measured is a global resolution of the detector, which include the effect of the non-uniformity of the detector all over the sensitive area. The 3 detectors are placed in parallel with mechanical method, the relative rotation of the 3 detectors will deteriorate the final measured resolution, which means that the measured resolution is much conservative.

\end{multicols}

\clearpage
\end{CJK*}

\end{document}